# Hybrid continuum-discrete macro-modelling of multi-ring masonry arch bridges

B. Pantò[1]*, C. Chisari[2], L. Macorini[1], B.A. Izzuddin[1]

[1] Department of Civil and Environmental Engineering, Imperial College London
South Kensington Campus, London SW7 2AZ, United Kingdom
[2] Department of Architecture and Industrial Design, University of Campania "Luigi Vanvitelli", Abazia di S. Lorenzo, via S. Lorenzo, Aversa CE (Italy)

## ABSTRACT

This paper presents a hybrid continuum-discrete macro-modelling strategy with a multiscale calibration procedure for realistic simulations of brick/block-masonry bridges. The response of these structures is affected by the intrinsic nonlinearity of the masonry material, which in turn depends upon the mechanical properties of units and mortar joints and the bond characteristics. Finite element approaches based upon homogenised representations are widely employed to assess the nonlinear behaviour up to collapse, as they are generally associated with a limited computational demand. However, such models require an accurate calibration of model material parameters to properly allow for masonry bond. According to the proposed approach, the macroscale material parameters are determined by an advanced multi-objective strategy with genetic algorithms from the results of mesoscale "virtual" tests through the minimisation of appropriate functionals of the scale transition error. The developed continuum-discrete finite element macroscale description and the calibration procedure are applied to simulate the nonlinear behaviour up to collapse of multi-ring arch-bridge specimens focusing on the 2D planar response. The results obtained are compared to those achieved using detailed mesoscale models confirming the effectiveness and accuracy of the proposed approach for realistic nonlinear simulations of masonry arch bridges.

**Keywords**: nonlinear analysis; continuous finite element models; mesoscale models; multi-ring masonry arches; optimisation procedures; genetic algorithms.

# 1 INTRODUCTION

Old masonry arch bridges belong to the cultural and architectural heritage and still play a critical role within railway and roadway networks in Europe and worldwide. These structures were built following empirical rules and were not designed to resist current traffic loading and the loads induced by extreme events, such as earthquakes. An accurate assessment of the ultimate performance of these complex structural systems represents a crucial step to prevent future failures and preserve such historical structures for the next generations.

Masonry arch barrels are the key structural components of masonry arch bridges. Their nonlinear behaviour is strongly influenced by the mechanical properties of the two constituents, masonry units and mortar joints, and their arrangement to form the brick/blockwork of the arch (i.e. masonry bond). Two main categories of masonry arch bridges can be identified: stone masonry and brick masonry bridges. In the first group, the arches are built from large voussoirs organised in a single arch ring [1]. Conversely, in the case of brick masonry bridges, a multi-ring arrangement is usually utilised, where the number of rings depends on the span length of the arch. The rings are typically bonded together using the stretcher method [2], where the connection between adjoining rings is guaranteed by continuous mortar joints. To date, numerous laboratory and in-situ tests have been performed to investigate the failure mechanisms of masonry arches and bridges, considering also the influence of backfill, under monotonic and cyclic loading conditions [3]. Specific studies on multi-ring arches showed how ring separation and shear sliding generally affect the ultimate strength and failure mode [4] [7], where weak circumferential mortar joints have been found to lead to an ultimate strength reduction of about 30% for short spans and up to 70% in the case of longer span arches.

In previous research, different numerical strategies have been proposed to simulate the nonlinear behaviour of masonry arches and bridges [3]. Generally, approaches based on limit analysis principles can be effectively used to estimate the ultimate load capacity [8]. However, such strategies do not provide information about the nonlinear response before collapse, and they are often based upon crude assumptions, e.g. the representation of masonry as a no-tension material, which may lead to underestimating the ultimate resistance of masonry arches. Previous studies also comprised simplified 2D finite element (FE) limit-analysis descriptions to simulate the arch-backfill interaction [9][10] and 3D nonlinear FE strategies with elasto-plastic solid elements [1][11][12], where masonry is assumed as a homogeneous isotropic material disregarding its anisotropic nature. These modelling approaches are widely employed in engineering practice due to their computational efficiency, especially for the analysis of large bridges. However, they generally disregard the anisotropic nature of masonry and require complex calibration procedures, as shown in [13] with reference to masonry walls.

More recent numerical models for masonry arched structures and bridges include the micro-model strategy proposed by Milani et al. [14] using triangular rigid elements and nonlinear links, the discrete macro-element method (DMEM) [15][16][17] and the distinct element method (DEM) [18][19]. A detailed 3D mesoscale modelling strategy for masonry arch bridges has been developed at Imperial College London [20][21], which is used as the reference solution for the calibration of the proposed macroscale approach hereinafter. According to this strategy, the masonry parts of the bridge are simulated by using linear solid elements and 2D nonlinear interface elements to explicitly allow for the masonry bond [22]. The backfill is modelled by elasto-plastic solid elements, and the connection between the masonry components and the backfill is represented through nonlinear interfaces allowing for the actual frictional interaction. This approach generally leads to accurate response predictions, including under extreme loading, but it is associated with significant computational cost which can hinder its use for the practical assessment of real large structures.

This paper proposes an efficient hybrid macroscale description for multi-ring masonry arches and masonry arches bridges. Elasto-plastic continuum solid elements interacting with 2D nonlinear interfaces are employed to model a masonry arch, though, unlike the mesoscale strategy, mesh discretisation is not directly related to the dimensions of units and mortar joints. The damage-plasticity model proposed in [13] and a multi-surface cohesive-frictional model [23] are employed for solid and interface elements, respectively. The mechanical parameters of the hybrid model are determined through a multi-objective optimisation procedure as in [13]. The proposed modelling and calibration strategies are validated considering two multi-ring masonry arch specimens characterised by different mechanical properties and failure modes, one of which allows for interaction with the backfill.

## 2  THE HYBRID MACRO-MODELLING APPROACH

In the proposed FE modelling strategy, the arch is discretised using a regular mesh of nonlinear continuum 20-noded solid elements. In addition, 2D nonlinear zero-thickness interface elements are arranged along the circumferential mid-thickness surface of the arch to simulate damage associated with potential ring sliding and separation. In the simplest case where only one circumferential layer of interfaces is considered (Figure 2), each interface lumps the linear deformability and non-linear behaviour of n-1 ring joints, with n being the number of rings of the physical arch. Importantly, the characteristics of the FE mesh with solid elements are not directly linked to the masonry bond. Thus, an arbitrary number of solid elements can be employed along the length of the arch, according to the desired level of response detail, but at least two solid elements should be arranged along the thickness of the arch to accommodate the mid-thickness nonlinear interfaces.

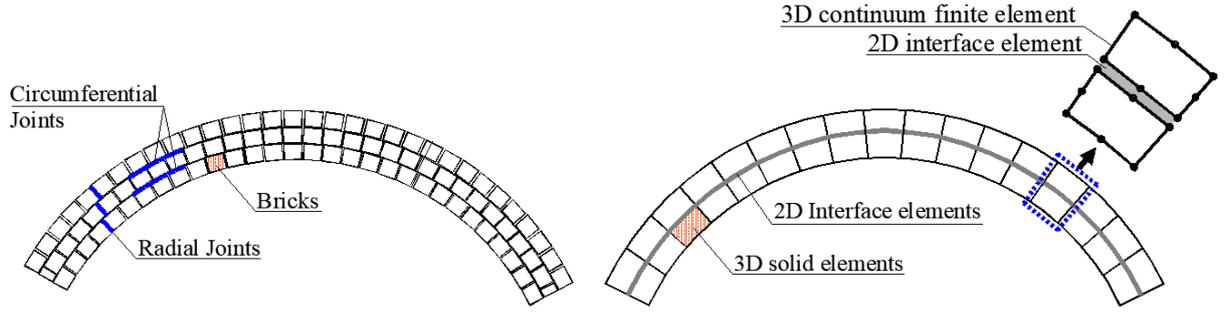

Figure 1. Generic multi-ring arch (a) and its macro-modelling description (b).

## 2.1 The 3D damage-plasticity model

In the macroscale representation implemented in ADAPTIC [26], the isotropic plastic-damage material model presented in [13] is used for the 20-noded solid elements. A standard decomposition of total strains ($\boldsymbol{\varepsilon}$) in elastic ($\boldsymbol{\varepsilon}_e$) and plastic ($\boldsymbol{\varepsilon}_p$) components is considered, and the stress tensor ($\boldsymbol{\sigma}$) is obtained from the effective stress tensor ($\bar{\boldsymbol{\sigma}}$) and a scalar damage variable $d(\bar{\boldsymbol{\sigma}}, \kappa_t, \kappa_c)$. The latter variable depends on the stress state and two historical variables ($\kappa_t, \kappa_c$) representing the evolution of plastic strains in tension and in compression. The material relationship is expressed analytically by:

$$\boldsymbol{\sigma} = (1-d)\,\bar{\boldsymbol{\sigma}} = (1-d)\,\boldsymbol{K_0}\,(\boldsymbol{\varepsilon} - \boldsymbol{\varepsilon}_p) \qquad (1)$$

where $\boldsymbol{K_0}$ is the initial fourth-order isotropic elastic tensor.

The local plastic problem is solved at each integration point of the domain to evaluate the effective stress, adopting a non-associated elasto-plastic constitutive law with Drucker-Prager-like plastic flow potential, according to the approach proposed in [27].

The plastic behaviour is governed by the evolution of the yield surface:

$$F(\bar{\sigma}, \kappa) = \frac{1}{1-\alpha} \cdot \left(\alpha I_1 + \sqrt{3J_2} + \beta(\kappa)\langle\bar{\sigma}_{max}\rangle - \gamma\langle-\bar{\sigma}_{max}\rangle\right) + \bar{f}_c(\kappa_c) \qquad (1)$$

where:

- $\beta(\kappa) = -\frac{\bar{f}_c(\kappa_c)}{\bar{f}_t(\kappa_t)}(1-\alpha) - (1+\alpha)$;

- $\alpha = \frac{\tilde{f}_{b0} - 1}{2\tilde{f}_{b0} - 1}$;

- $\gamma = \frac{3(1-K_c)}{2K_c - 1}$;

- $\bar{\sigma}_{max} = \max(\bar{\sigma}_1, \bar{\sigma}_2, \bar{\sigma}_3)$ with $\bar{\sigma}_i$ principal effective stress;

- $\langle x \rangle = \frac{x + |x|}{2}$.

- $\bar{f}_c(\kappa_c), \bar{f}_t(\kappa_t)$ effective strength in compression and tension, respectively;

- $K_c$ ratio of the second stress invariant on the tensile meridian to that on the compressive meridian at initial yield;

- $\tilde{f}_{b0}$ ratio between biaxial and uniaxial compressive strength.

To improve the computational robustness, both tensile and compressive strengths, $\bar{f}_\chi(\kappa_\chi)$ with $\chi = t, c$, allow for hardening behaviour, while the softening response is obtained for the nominal strength $f_\chi(\kappa_\chi)$ by introducing an appropriate damage law $d_\chi(\kappa_\chi) = 1 - \frac{f_\chi(\kappa_\chi)}{\bar{f}_\chi(\kappa_\chi)}$, as shown in Figure 2.

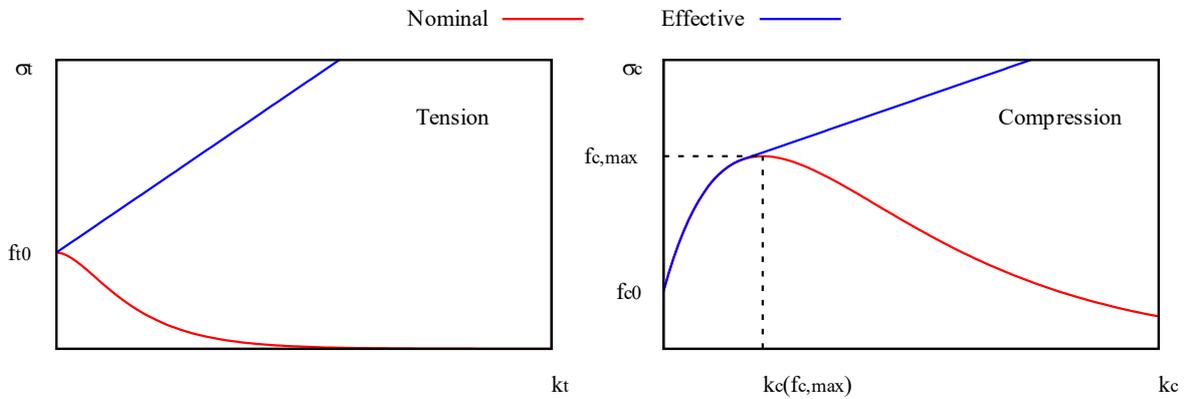

Figure 2. Uniaxial constitutive relationships in tension and compression.

The global damage variable is obtained as:

$$d(\bar{\boldsymbol{\sigma}}, \boldsymbol{\kappa}) = 1 - [1 - s_t(\bar{\boldsymbol{\sigma}}) \, d_c(\kappa_c)][1 - s_c(\bar{\boldsymbol{\sigma}}) \, d_t(\kappa_t)] \qquad (3)$$

where:

- $s_t(\bar{\boldsymbol{\sigma}}) = 1 - w_t r(\bar{\boldsymbol{\sigma}})$;

- $s_c(\bar{\boldsymbol{\sigma}}) = 1 - w_c\big(1 - r(\bar{\boldsymbol{\sigma}})\big)$;

- $r(\bar{\boldsymbol{\sigma}}) = \begin{cases} 0 & if\ \bar{\sigma}_1 = \bar{\sigma}_2 = \bar{\sigma}_3 = 0 \\ \frac{\sum_{i=1}^{3} \langle \bar{\sigma}_i \rangle}{\sum_{i=1}^{3} |\bar{\sigma}_i|} & otherwise \end{cases}$, scalar parameter ranging from 0 (all principal stresses are negative) to 1 (all principal stresses are positive) expressing the state stress;

- $w_t, w_c$ are parameters governing the stiffness recovery from compression to tension and vice versa.

The model has been extensively used to simulate the mechanical behaviour of concrete [27][33] and masonry [36][37][38]. Some inherent model characteristics, however, hinder its use to represent specific shear failure modes typical of multi-ring masonry arches. More specifically, the adopted damage-plasticity continuum description does not enable the definition of the shear strength independently from the tension and compression strengths. It can be seen by applying Eq. (1) assuming a pure shear 2D stress state ($\bar{\sigma}_x = \bar{\sigma}_y = \bar{\sigma}_z = \bar{\tau}_{xy} = \bar{\tau}_{yz} = 0, \bar{\tau}_{xz} = \bar{\tau}$) which leads to the yield function:

$$F(\bar{\tau}, \boldsymbol{\kappa}) = \frac{1}{1 - \alpha} \cdot \left(\sqrt{3}\bar{\tau} + \beta(\boldsymbol{\kappa})\bar{\tau}\right) + \bar{f}_c(\kappa_c) = 0 \qquad (4)$$

Imposing $F(\bar{\tau}, \boldsymbol{\kappa}) = 0$ in Eq. (4), the effective shear strength $\bar{f}_v(\boldsymbol{\kappa})$ can be evaluated as:

$$\bar{f}_v(\boldsymbol{\kappa}) = \frac{1-\alpha}{\sqrt{3}+\beta(\boldsymbol{\kappa})}|\bar{f}_c(\kappa_c)| \tag{5}$$

and, since $r(\bar{\boldsymbol{\sigma}}) = 0.5$ for pure shear, the damage parameter becomes:

$$d(\boldsymbol{\kappa}) = 1 - [1 - (1 - 0.5\, w_t)\, d_c(\kappa_c)][1 - (1 - 0.5\, w_c)\, d_t(\kappa_t)] \tag{6}$$

Assuming that damage in compression has not developed, $d_c(\kappa_c) = 0$, the equivalent damage parameter becomes:

$$d(\boldsymbol{\kappa}) = (1 - 0.5\, w_c)\, d_t(\kappa_t) \tag{7}$$

and

$$\begin{aligned}f_v(\boldsymbol{\kappa}) &= (1 - d(\boldsymbol{\kappa}))\bar{f}_v(\boldsymbol{\kappa}) \\ &= [1 - (1 - 0.5\, w_c)\, d_t(\kappa_t)]\frac{1-\alpha}{\sqrt{3}+\beta(\boldsymbol{\kappa})}|\bar{f}_c(\kappa_c)|\end{aligned} \tag{8}$$

From Eq. (8), some considerations can be made on the shear behaviour of the model. Firstly, the initial shear strength (Eq. (5) with $\boldsymbol{\kappa} = \boldsymbol{0}$) is governed by the initial compression and tension strengths and the parameter $\tilde{f}_{b0}$, which in practice is always in the range $1.12 \div 1.16$. This confirms that it is not possible to define a specific shear strength independent from tension and compression strengths, as for instance a shear strength relating to the sliding of mortar joints, which is a typical shear failure mode for multi-ring masonry arches. A workaround to have some freedom in the definition of initial shear strength could be to calibrate $\bar{f}_c(0) = f_{c0}$ appropriately and independently from the observed compressive behaviour, while both $f_{t0}$ and $f_{c,max}$ would still be determined based on their specific failure modes.

The second consideration is that the evolution of nominal shear strength depends on the parameter $w_c$ (see Eq. (8)) which is defined based on the expected cyclic response (stiffness recovery). A typical value, leading to complete stiffness recovery from tension to compression, is $w_c = 1.0$ [34][35]. Inserting this in Eq. (7), the expression for damage in pure shear in the absence of compression damage is obtained $d(\kappa_t) = 0.5\, d_t(\kappa_t)$. The conclusion is that the evolution of nominal shear strength is completely governed by damage in tension, without the possibility for specifying an alternative more realistic constitutive relationship.

Finally, it is worth mentioning that the macroscale damage-plasticity continuum representation is not capable of distinguishing failure due to shear parallel to the mortar bed joint $\tau_{zx}$ from that due to shear orthogonal to the mortar bed joint $\tau_{xz}$, as in the Cauchy solid these two stresses are equal, and the yield surface cannot consider separate contributions. In reality, while the former failure mode is governed by sliding of the units on the weak planes represented by the mortar joints, the latter is governed by the internal rotation of bricks depending on their geometric shape ratio and brick interlocking, as schematically shown in **Error! Reference source not found.** (Casolo, 2004). To allow for these different phenomena enriched continuum representations, e.g. Cosserat continuum [28], would need to be employed.

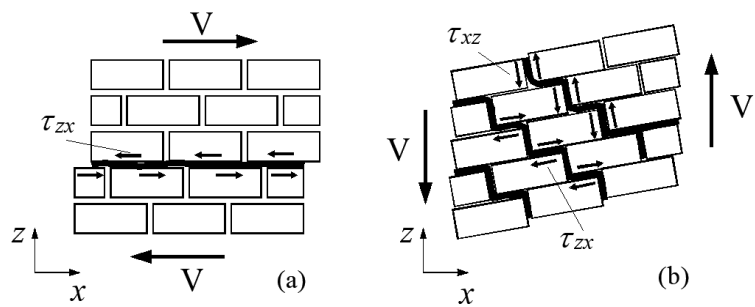

Figure 3.    Shear stress and failure mode in a brick-work masonry under pure shear load parallel (a) and orthogonal (b) to the bed joints.

To overcome these intrinsic limitations of typical continuum damage-plasticity constitutive models, an alternative hybrid macroscale representation is proposed, in which, as outlined before, shear sliding along the continuous circumferential mortar joints of multi-ring arches is described by introducing nonlinear interfaces whose material characteristics are defined based on the calibration strategy described in Section 3.

## 2.2  Constitutive model for nonlinear interface elements

2D 16-noded interface elements [22] are employed for the mid-thickness circumferential interfaces using the plasticity-damage constitutive model proposed in [23]. According to this description, the stress and strain fields are composed of a normal component in the direction orthogonal to the interface and two shear components on the plane of the interface. The effective stresses are evaluated at each Gauss point by solving a linear hardening elasto-plastic problem considering multi-surface plasticity. Then, the nominal stresses are obtained by multiplying the effective stresses by the damage matrix $\boldsymbol{D}$, containing the damage index in tension, shear and compression ranging from 0 (no-damage) to 1 (complete damage).

Similarly to the solid elements, a standard decomposition between elastic and plastic deformations is considered and the concept of effective stress $\bar{\boldsymbol{\sigma}} = \mathbf{K}_0(\boldsymbol{\varepsilon} - \boldsymbol{\varepsilon}_p)$ is introduced, where $\mathbf{K}_0 = diag\{k_n \quad k_n \quad k_n\}$ is the diagonal initial stiffness matrix with $k_n$ and $k_t$ the normal and shear stiffness, $\bar{\boldsymbol{\sigma}} = [\bar{\sigma} \ \bar{\tau}_1 \ \bar{\tau}_2]$, $\boldsymbol{\varepsilon} = [\varepsilon \ \gamma_1 \ \gamma_2]$ and $\boldsymbol{\varepsilon}_p = [\varepsilon_p \ \gamma_{p1} \ \gamma_{p2}]$ the effective stress, the total strains and the plastic strains, respectively. The nominal stresses are evaluated from the effective stress according to:

$$\boldsymbol{\sigma} = (\mathbf{I}_3 - \mathbf{D})\bar{\boldsymbol{\sigma}} = (\mathbf{I}_3 - \mathbf{D})\mathbf{K}_0(\boldsymbol{\varepsilon} - \boldsymbol{\varepsilon}_p) \qquad (9)$$

where $\mathbf{D}$ represents an anisotropic damage tensor, containing distinct variables for the normal ($D_n$) and the tangential ($D_t$) directions. A tri-linear plastic yield domain is considered to simulate the tensile (Mode I), shear (Mode II) and crushing (Model III) failure models. Three distinct plastic works, corresponding to each fracture mode rule the evolution of the damage variables.

The plastic yield domain (Figure 4) is composed of three surfaces, $F_t$, $F_c$, and $F_s$ respectively, associated with the tensile (mode I), compression and shear (mode II) failure modes, as defined by:

$$F_s(\overline{\boldsymbol{\sigma}}, q) = \sqrt{\overline{\tau_1^2} + \overline{\tau_2^2}} + \tilde{\sigma} \tan(\phi) - c' \tag{10a}$$

$$F_t(\overline{\boldsymbol{\sigma}}, q) = \overline{\sigma} - (f_t - q) \tag{10b}$$

$$F_c(\overline{\boldsymbol{\sigma}}) = -\overline{\sigma} + f_c \tag{10c}$$

where $f_t$ and $f_c$ are the tensile and compression material strengths, $\phi$ the friction angle and $q$ a linear hardening variable, ranging from 0 (initial value) to the limit value $q_{lim} = \frac{c}{\tan}(\phi) - f_t$. Moreover, $c' = c$ if $q \leq q_{lim}$ and $c' = c + (q - q_{lim})\tan(\phi)$ if $q > q_{lim}$. With the increase of $q$ the surface $F_t$ reduces until becoming a point when $q$ reaches the value $q_{lim}$. On the other hand, $F_s$ increases with the increase of $q$ (Figure 4). Two associated plastic flows are related to $F_t$ and $F_c$, while a plastic potential $G_s$ in shear, obtained from $F_s$ substituting $\phi$ to $\phi_g$, is considered to take into account the effects of masonry dilatancy.

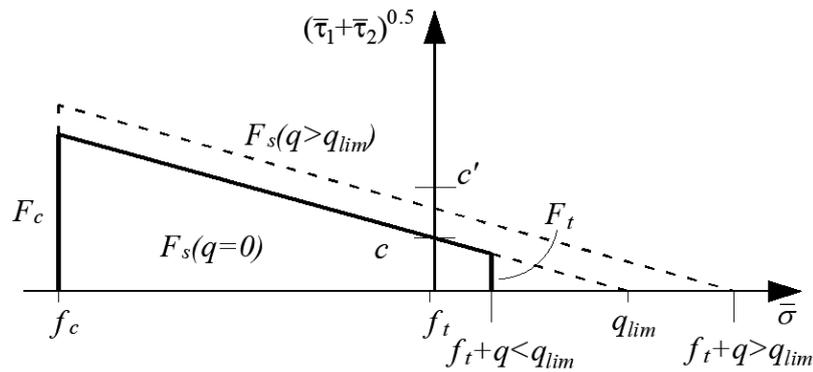

Figure 4. Yield surface (a) and qualitative yield function (b) of the interface.

Following the solution of the plastic problem, the damage evolution is evaluated as a function of the three ratios $r_c = W_{pc}/G_c$, $r_t = W_{pt}/G_t$ and $r_s = W_{ps}/G_s$ where $W_{pc}, W_{pt}$ and $W_{ps}$ are the plastic works in compression, tension and shear, respectively, and $G_c, G_t, G_s$ the corresponding fracture energies. Finally, the nominal stresses are given by Eq. (8). More details on the model formulation can be found in [23].

## 3   CALIBRATION PROCEDURE

The mechanical calibration of the proposed model requires the determination of several material parameters defining the linear and nonlinear behaviour of the 3D solid elements and the 2D interfaces, as described Sections 2.1 and 2.2, and reported in [22] and [23]. For this reason, an objective and robust calibration procedure represents a fundamental step to guarantee the model accuracy and applicability.

In this work, a modified version of the multiscale calibration procedure proposed in [13] is applied. This is based on the representation of a structure under suitable boundary conditions according to two scales: mesoscale, indicated hereinafter by the superscript *m*, and macroscale, with the superscript *M*. The considered setup is called *virtual test*, and it is assumed there exists a mapping $\mathcal{M}: \Omega^m \to \Omega^M$ between the mesoscale and the macroscale domains. Stress power equivalence between the two scales is approximately enforced on the entire domain of the virtual test. The stress power equivalence reads:

$$\int_{\Omega^M} \boldsymbol{\sigma}^M : \dot{\boldsymbol{\varepsilon}}^M d\Omega^M = \int_{\Omega^m} \boldsymbol{\sigma}^m : \dot{\boldsymbol{\varepsilon}}^m d\Omega^m + \dot{\epsilon} \qquad (10)$$

where $\dot{\epsilon}$ represents the error rate due to the approximations induced by the specific macromodel utilised. Considering pseudo-static stress states, the equality between internal and external work implies:

$$\int_{\Gamma^M} \boldsymbol{t}^M \cdot \dot{\boldsymbol{u}}^M d\Gamma^M + \int_{\Omega^M} \boldsymbol{b}^M \cdot \dot{\boldsymbol{u}}^M d\Omega^M$$
$$= \int_{\Gamma^m} \boldsymbol{t}^m \cdot \dot{\boldsymbol{u}}^m d\Gamma^m + \int_{\Omega^m} \boldsymbol{b}^m \cdot \dot{\boldsymbol{u}}^m d\Omega^m + \dot{\epsilon} \qquad (11)$$

where $\boldsymbol{t}$ are the surface forces on the boundary $\Gamma$, while $\boldsymbol{b}$ are volume forces. Neglecting the contribution of these latter and considering the chain rule of differentiation, Eq. (11) finally reads:

$$\dot{\epsilon} = \int_{\Gamma^M} \left( \boldsymbol{t}^M \cdot \dot{\boldsymbol{u}}^M - \boldsymbol{t}^m \cdot \dot{\boldsymbol{u}}^m \frac{\partial \Gamma_i^m}{\partial \Gamma_i^M} \right) d\Gamma_i^M \qquad (12)$$

Eq. (12) represents the error rate at time $t$ due to the scale transition. In [13], a global non-negative monotonically increasing error function was defined:

$$\epsilon(t) = \int_0^t [\dot{\epsilon}(\tau)]^2 d\tau$$
$$= \int_0^t \left[ \int_{\Gamma^M} \left( \boldsymbol{t}^M(\tau) \cdot \dot{\boldsymbol{u}}^M(\tau) - \boldsymbol{t}^m(\tau) \right. \right. \qquad (13)$$
$$\left. \left. \cdot \dot{\boldsymbol{u}}^m(\tau) \frac{\partial \Gamma_i^m}{\partial \Gamma_i^M} \right) d\Gamma_i^M \right]^2 d\tau$$

However, it is also possible to partition the error defined as in Eq. (10) or in Eq. (12):

$$\dot{\epsilon} = \dot{\epsilon}_1 + \dot{\epsilon}_2 + \cdots$$

$$= \int_{\Omega_1^M} (\boldsymbol{\sigma}^M : \dot{\boldsymbol{\varepsilon}}^M - \boldsymbol{\sigma}^m : \dot{\boldsymbol{\varepsilon}}^m) d\Omega_1^M \quad (14)$$

$$+ \int_{\Omega_2^M} (\boldsymbol{\sigma}^M : \dot{\boldsymbol{\varepsilon}}^M - \boldsymbol{\sigma}^m : \dot{\boldsymbol{\varepsilon}}^m) d\Omega_2^M + \cdots$$

$$\dot{\epsilon} = \dot{\epsilon}_1 + \dot{\epsilon}_2 + \cdots$$

$$= \int_{\Gamma_1^M} (\boldsymbol{t}^M \cdot \dot{\boldsymbol{u}}^M - \boldsymbol{t}^m \cdot \dot{\boldsymbol{u}}^m) d\Gamma_1^M \quad (2)$$

$$+ \int_{\Gamma_2^M} (\boldsymbol{t}^M \cdot \dot{\boldsymbol{u}}^M - \boldsymbol{t}^m \cdot \dot{\boldsymbol{u}}^m) d\Gamma_2^M + \cdots$$

The contributions $\dot{\epsilon}_1, \dot{\epsilon}_2, \ldots$ refer to a volume partitioning in Eq. (14), with $\Omega_1^{M|m} + \Omega_2^{M|m} + \cdots = \Omega^{M|m}$, or load-based partitioning in Eq. (2). For the sake of simplicity, in Eq. (14), (2) it is assumed that there is not any modification of volumes and surfaces in the scale transition, i.e., $\partial \Gamma_i^m / \partial \Gamma_i^M = \partial \Omega_i^m / \partial \Omega_i^M = 1$.

In this case several error functions can be defined as:

$$\omega_i = \int_0^T [\dot{\epsilon}_i(\tau)]^2 d\tau \quad i = 1,2,\ldots \quad (3)$$

The solution of the calibration procedure is given by the solution of the multi-objective minimisation problem:

$$\widetilde{\boldsymbol{p}} = \arg\min_{\boldsymbol{p}} [\omega_1, \omega_2, \ldots] \quad (4)$$

The error partitioning defined in Eq. (14) or (2) has two consequences. The first consequence is that it allows defining the granularity of the homogenisation, avoiding the possible error compensations given by different parts of the structure. For instance, if the nonlinearities in the mesoscale model are concentrated in one small region of the domain, it is possible to use volume partitioning in Eq. (14) to focus the calibration of the parameters governing the nonlinear behaviour of the macroscale representation in that region, while controlling the elastic parameters by matching the response in the remaining domain. The second consequence is that the calibration problem is turned into a multi-objective optimisation problem. As shown in [29] and [39], using multiple objectives in a calibration problem may strongly increase the robustness of the procedure.

The multi-objective optimisation problem will be solved by means of a Non-Dominated Sorting Genetic Algorithm [24], implemented in TOSCA-TS software [25]. The optimum is represented by the Pareto Front (PF), which represents the set of non-dominated solutions. A careful investigation on the features of the Pareto Front may also highlight possible inconsistencies of the model to calibrate [29].

Finally, it is worth noting that the procedure considers the evolution of the stress power over time, and thus it cannot properly allow for the additional work contribution due to initial loading. Thus, it is preferable to avoid initial loading in the virtual test. However, this does not limit the applicability of the procedure, as multiple loads with independent loading paths can be introduced without any modifications in the methodology.

## 4  NUMERICAL EXAMPLES

Two masonry arch specimens, one interacting with backfill as found in typical masonry arch bridges, are analysed by means of detailed mesoscale models [22]. The results of the mesoscale analyses are then used as reference solutions to highlight some limits of a typical continuum macroscale description, and to investigate the improved accuracy guaranteed by the proposed continuum-discrete hybrid representation for multi-ring masonry arches.

### 4.1 Masonry arch and bridge specimens

The first specimen (Figure 5a) consists of a 5 m span three-ring brick-masonry arch. The arch is characterised by 1250 mm rise, 330 mm thickness and 675 mm width. Adjacent rings with $215 \times 102.5 \times 65$ mm$^3$ bricks are connected according to the stretcher method by continuous circumferential 10 mm thick mortar joints. The second specimen (confined arch) comprises a brick-masonry arch with the same geometrical characteristics of the bare arch interacting with backfill material, which extends 2460 mm horizontally from the two supports of the arch and 300 mm vertically above the crown (Figure 5b). Full supports are assumed at the base of the arch and the backfill, while simple supports against the horizontal longitudinal displacements are applied on the two vertical sides of the backfill. Moreover, the horizontal transverse displacements on the two lateral faces of the arch and backfill are restrained to represent a plane strain condition (Figure 5b).

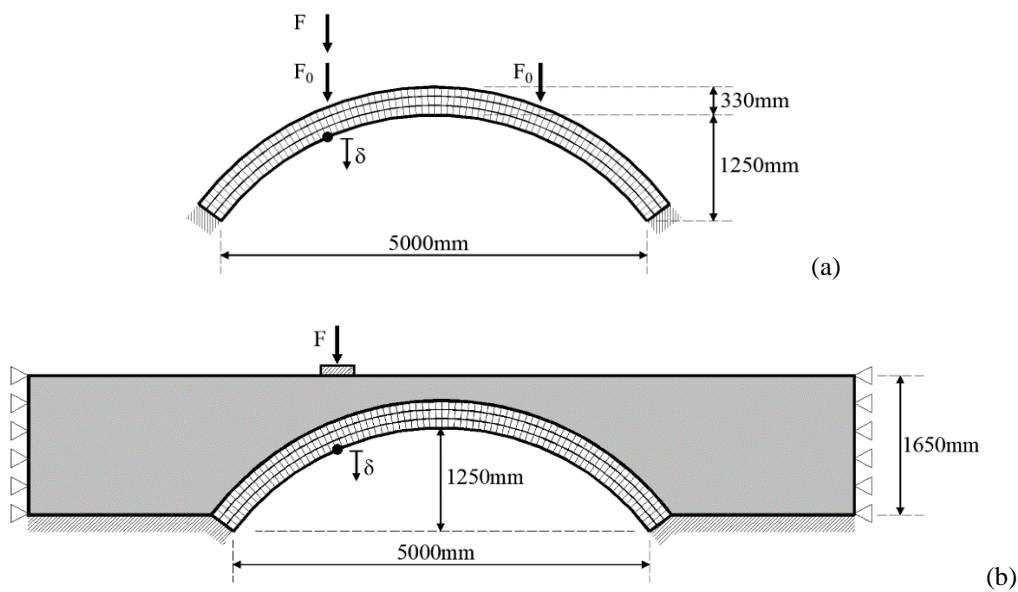

Figure 5. Geometrical characteristics and loading conditions for the bare (a) and confined (b) arch specimens.

## 4.2 Mesoscale simulations

In the numerical mesoscale description, 20-noded elastic solid elements are used to simulate the brick units and 16-noded interfaces [22] are employed to represent both the radial and the circumferential mortar joints. The mesoscale description of the arch requires 240 3D solid elements, 403 2D interface elements and 6453 nodes to which correspond 19359 DOFs. The backfill is modelled adopting a FE mesh with 15-noded tetrahedral elements. Finally, nonlinear interface elements are utilised to model the physical interface connecting the arch to the backfill.

Two masonry types have been considered in the analyses: a *strong* masonry to represent modern good quality brickwork, and a *weak* masonry to represent historical masonry [30]. The mesoscale mechanical parameters are reported in Tables 1 and 2. They were considered also in previous numerical studies [20] [31], where the adopted mesoscale description for masonry arches and bridges was validated against experimental tests.

Following [21], an elasto-plastic material model with a modified Drucker-Prager yield criterion is employed for the backfill, assuming a Young's modulus $E_b = 500 MPa$, a cohesion $c_b = 0.001$MPa, a friction and a dilatancy coefficient $tan\phi_b = 0.95$ and $tan\psi_b = 0.45$. The nonlinear interfaces simulating the interaction between the arch and the backfill at the extrados of the arch have tensile strength $f_{fi} = 0.002 MPa$, cohesion $c_{fi} = 0.0029 MPa$, friction coefficient $tan\phi_{fi} = 0.6$ and zero dilatancy.

Table 1. Mechanical parameters of the bricks adopted in the analyses.

| Masonry | $E_b$ [MPa] | $\nu$ [−] | $w$ [kN/m3] |
|---|---|---|---|
| Weak | 6000 | 0.15 | 16 |
| Strong | 16000 | 0.15 | 22 |

Table 2. Interface mechanical parameters adopted in the analyses.

| Masonry | $k_n - k_t$ [N/mm3] | $f_t - f_c - c$ [MPa] | $G_t - G_s - G_c$ [N/mm] | $tg\phi - tg\phi_g$ |
|---|---|---|---|---|
| Weak | 60.0 - 30.0 | 0.05 - 9.1 - 0.085 | 0.02 - 0.125 - 5.0 | 0.5 - 0.0 |
| Strong | 90.0 - 40.0 | 0.26 - 24.5 - 0.40 | 0.12 - 0.125 - 5.0 | 0.5 - 0.0 |

In the numerical simulations of the bare arch, two initial vertical forces $F_0$ = 22.5kN are applied at the quarter and three-quarter span and then maintained constant during the subsequent loading stage, when a vertical force $F$ is applied at quarter span and monotonically increased up to collapse. Both forces $F_0$ and $F$ are uniformly distributed on a patch area of 210×675mm². When the arch interacts with the backfill, the initial load corresponds to the weight of the arch and the backfill both with a specific weight of 22kN/m³, while the force $F$ is applied on the top surface of the backfill on a patch area of 400×675mm² centred at the quarter span of the arch (Figure 5b). To improve the numerical stability, nonlinear dynamic analysis is performed by imposing an initial velocity of 0.1mm/s at the loaded nodes, which is maintained constant during the simulation up to collapse. Zero viscous damping is considered in the analyses.

Figure 6 shows the load-displacement curves of the bare arch (Figure 6a) and the arch interacting with backfill (Figure 6b), where the force $F$ is plotted against the vertical deflection at the quarter span of the arch. A significant influence of the masonry typology on the global response is observed both in the case of the bare arch, where the ultimate load ranges from 31kN to 66kN, and for the arch with backfill, where the peak force varies from 74kN to 137kN. As expected, the initial stiffness of the bare arch is significantly affected by the masonry characteristics. Conversely, the confined arch shows almost the same initial stiffness for weak and strong masonry.

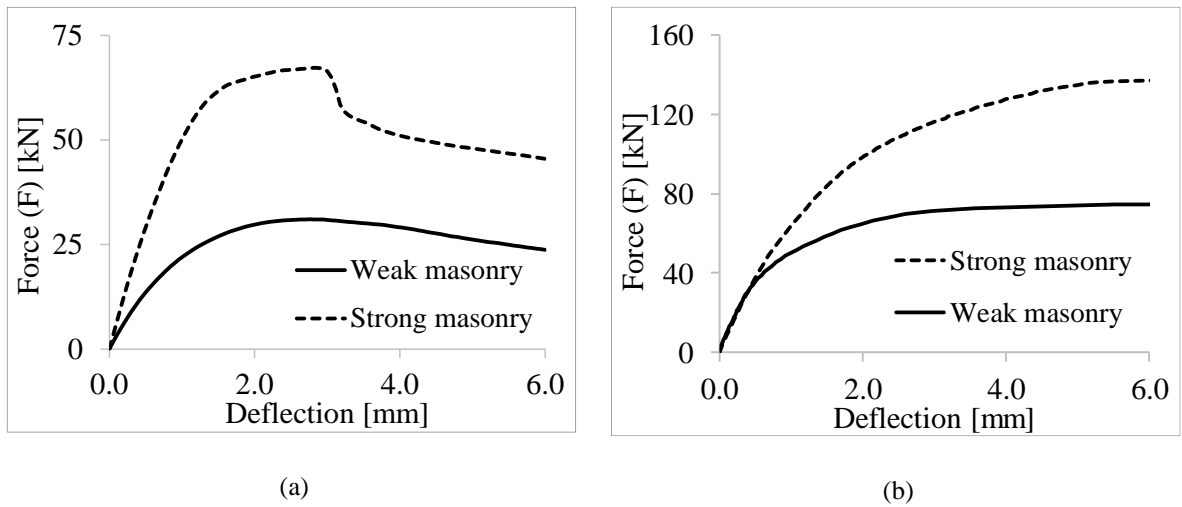

Figure 6. Load-deflection curves for the bare arch (a) and the arch interacting with backfill (b).

The failure mechanisms and the equivalent von-Mises stress contours of the two specimens, with both masonry typologies, are reported in Figure 7. Finally, Figure 8 shows the tensile damage contours at the last step of the analysis obtained by the different models. The failure mechanism of the models with weak masonry is characterised by shear sliding along the circumferential interfaces, mainly concentrated in the zone between the left support of the arch and the loading area at quarter span, and close to the three-quarter span of the arch. This mechanism prevents the activation of flexural plastic hinges. In the case of strong masonry, flexural failure is observed with the opening of four radial cracks in both the bare arch (Figure 7) and the arch with backfill (Figure 7d). In the models with weak masonry, significant damage in the radial joints is observed close to the load. In the models with backfill, large portions at the extrados of the arch are affected by shear-sliding damage, both at the radial and circumferential interfaces. This damage develops also at the frame-backfill interfaces.

In the following, the mesoscale solutions are assumed as the baseline results for the assessment of more efficient macroscale models and the proposed hybrid continuous-discrete descriptions for multi-ring arches.

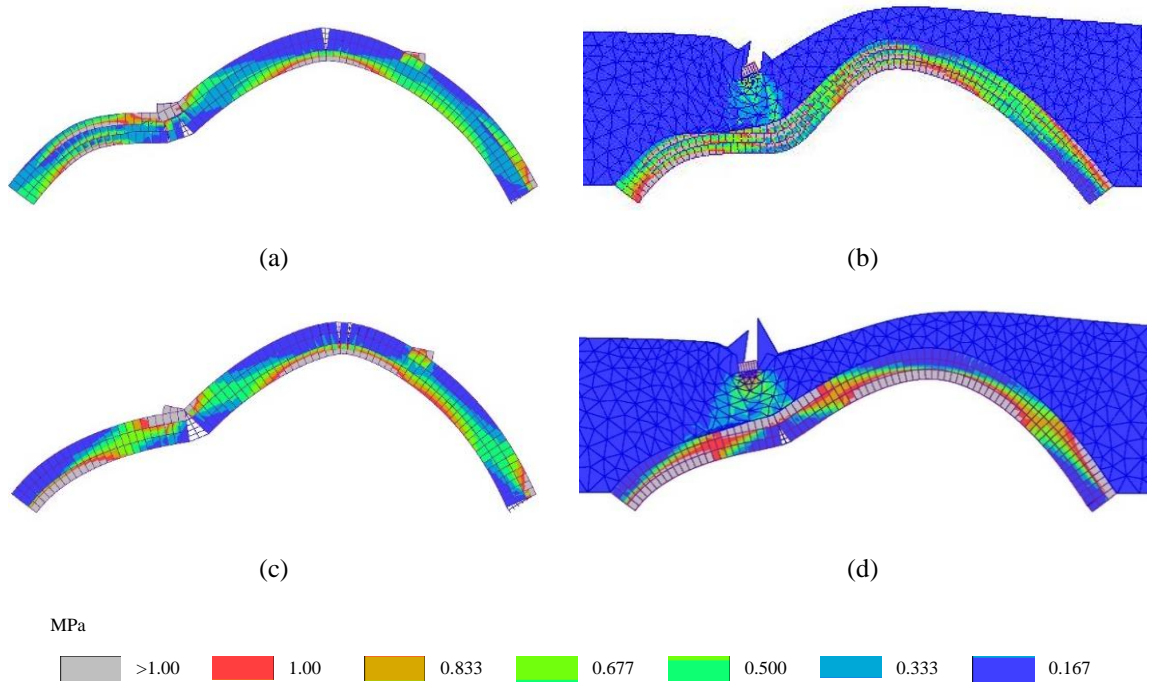

Figure 7. Ultimate deformed shape and the Von-Mises stress patterns: bare arch with (a) weak masonry and (c) strong masonry; arch interacting with backfill with (b) weak masonry and (d) strong masonry.

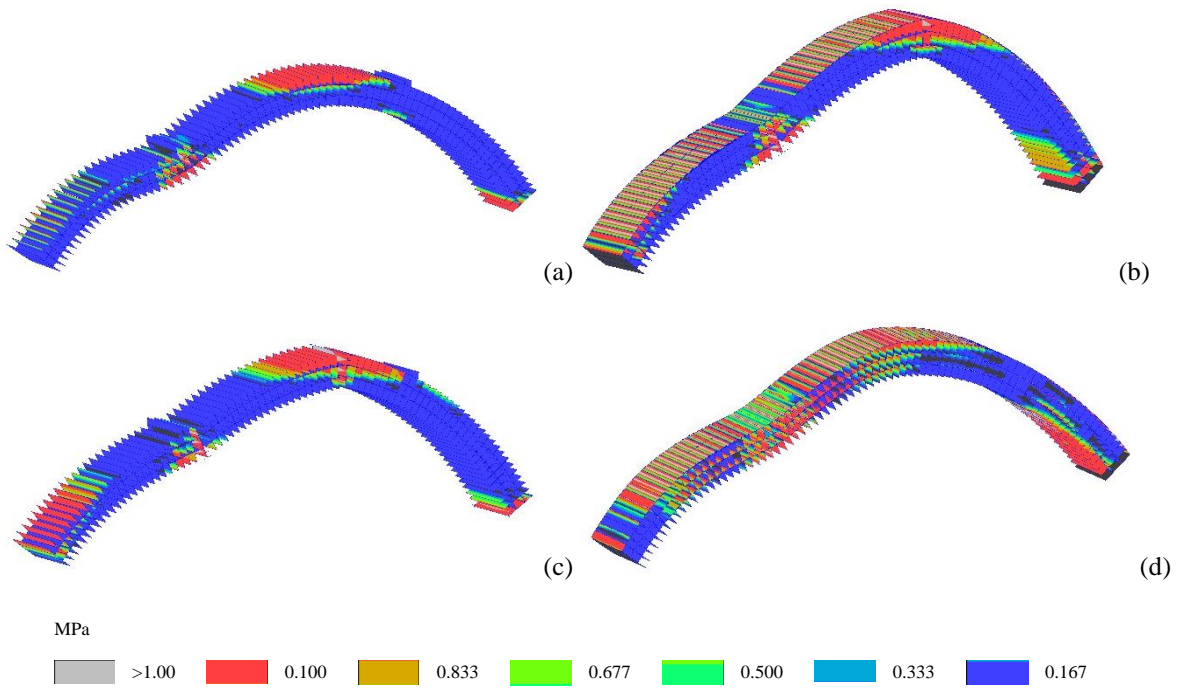

Figure 8. Interface tensile damage: bare arch with (a) weak masonry and (c) strong masonry; arch interacting with backfill with (b) weak masonry and (d) strong masonry.

## 4.3 Macroscale simulations

The two specimens presented in Section 4.1 have been analysed by a continuum macroscale description for masonry, using the isotropic damage-plasticity constitutive law described before. In the models, the masonry arch is represented by 80 3D solid elements, 40 2D interface elements and 981 nodes corresponding to 2943 DOFs. It can be observed that the macro-modelling description allows a reduction of 95% of DOFs compared to the mesoscale description demonstrating the potential for considerable reduction in computational demands with the proposed model.

The aim of this investigation is to explore the accuracy and potential limitations of a standard continuum isotropic macroscale approach to predict the response of multi-ring masonry arches, where the model material parameters are calibrated according to two alternative *simplified* and *advanced* procedures.

*4.3.1. Simplified calibration procedure*

In initial macroscale simulations, the material model parameters for masonry have been evaluated through a simplified calibration procedure considering the mesoscale material properties reported in Tables 1 and 2. The macroscopic Young's modulus for the masonry material $E$ has been determined by combining in series the stiffness of brick units with that of the mortar interfaces along the direction of the arch. The tensile strength and fracture energy $f_t, G_t$ and the compressive strength $f_c$ are assumed coincident to the corresponding values of the mesoscale interfaces. The remaining parameters for the damage plasticity model are assumed equal to standard values used in previous studies for modelling masonry materials. More specifically:

- The ratio between initial and maximum compressive strength $\tilde{f}_y = \frac{f_{c0}}{f_{c,max}}$ is assumed equal to 0.3 according to [27][32][33];

- The dilatancy angle $\psi$ is taken equal to 35° which is consistent with the value adopted for modelling quasi-brittle material as concrete [27][33] and corresponds approximately to the median of the values (ranging from 10° to 50°) typically used for masonry [36][37][38];

- The eccentricity of the plastic flow potential is taken as $\epsilon = 0.1$ to improve computational robustness as suggested in [34];

- $\mu$ governing the relative influence of damage and plasticity in tension ($\mu = 0$ for fully damage material) is assumed equal to 0.2;

- The plastic strain at maximum compression stress $k_{c,fc}$ is taken as 0.002 following [32];

- The ratio between the plastic strain at damage onset in compression and the plastic strain at maximum compression $\rho_c$ is considered equal to 1.0, as damage is assumed to develop in the softening branch of the stress-strain response.

As noted in Section 2.1, preliminary numerical simulations showed a significant influence of the parameter governing the stiffness recovery in compression $w_c$ on the global response of the arch. Thus, two limit values (0,1) are considered, while parameter $w_t$ determining the stiffness recovery in tension is assumed as zero. The complete set of mechanical properties for the continuum macro-modelling description are reported in Table 3.

Table 3. Macroscopic mechanical parameters resulting from the.

| $E$ | $\nu$ | $\tilde{f}_{b0}$ | $\tilde{f}_y$ | $\psi$ | $\epsilon$ | $K_c$ | $f_{mt}$ | $f_{mc}$ | $G_{mt}$ | $\mu$ | $k_{c,fc}$ | $\rho_c$ | $w_c$ | $w_t$ |
|---|---|---|---|---|---|---|---|---|---|---|---|---|---|---|
| [MPa] | [-] | [-] | [-] | [°] | [-] | [-] | [MPa] | [MPa] | [N/mm] | [-] | [-] | [-] | [-] | [-] |
| weak masonry | | | | | | | | | | | | | | |
| 2571 | 0.15 | 1.16 | 0.3 | 35 | 0.1 | 0.66 | 0.05 | 9.1 | 0.02 | 0.2 | 2E-3 | 1.0 | 0.0 1.0 | 0.0 |
| strong masonry | | | | | | | | | | | | | | |
| 4747 | 0.15 | 1.16 | 0.3 | 35 | 0.1 | 0.66 | 0.26 | 24.0 | 0.12 | 0.2 | 2E-3 | 1.0 | 0.0 1.0 | 0.0 |

Figure 9 shows the load-displacement responses predicted by the macroscale descriptions which are compared against the reference mesoscale curves. In the main, the macromodels predict the initial stiffness of the masonry arches accurately, yet significantly overestimating the peak strengths without providing a realistic representation of the post-peak behaviour as given by the reference mesoscale models. Furthermore, very different macromodel curves are obtained depending on the adopted value for $w_c$. In particular, the largest differences between the mesoscale models and the corresponding macromodels are achieved when $w_c$=1. It should be noted that this value is recommended by most software implementations [34][35] to model the cyclic response of quasi-brittle materials.

In Figure 10, the influence of the dilatancy angle on the global response of the weak masonry bare arch is shown. Since this parameter governs the normal plastic deformation due to shear, it is expected that by increasing ψ the global behaviour becomes more ductile due to the confinement effects exerted by the surrounding elements. Given its high influence on the global behaviour, it is apparent that more accurate calibration is needed for such critical parameter.

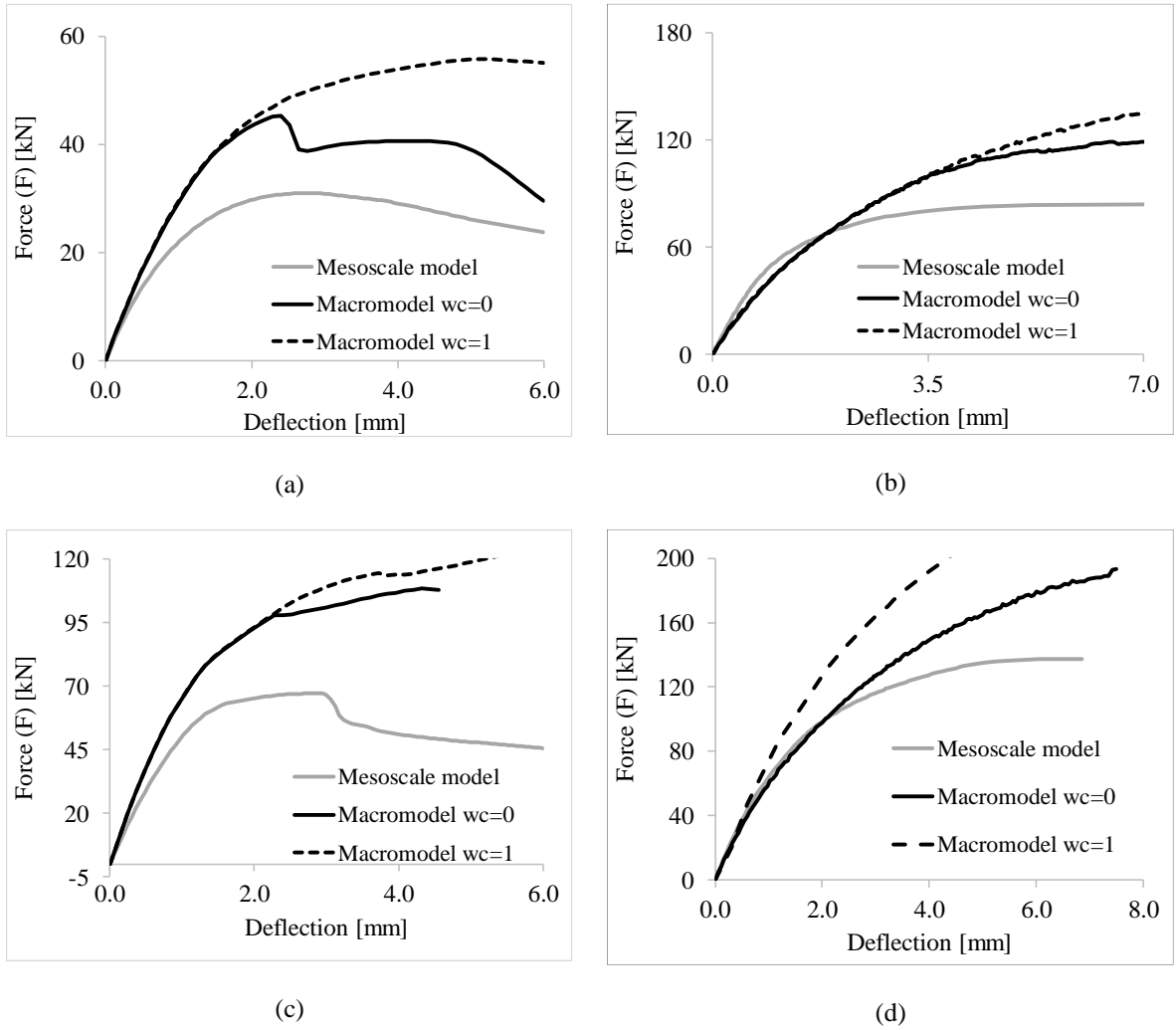

Figure 9. Load-displacement curves of (a) the bare arch and (b) the confined arch with weak masonry and (c) the bare arch and (d) confined arch with strong masonry.

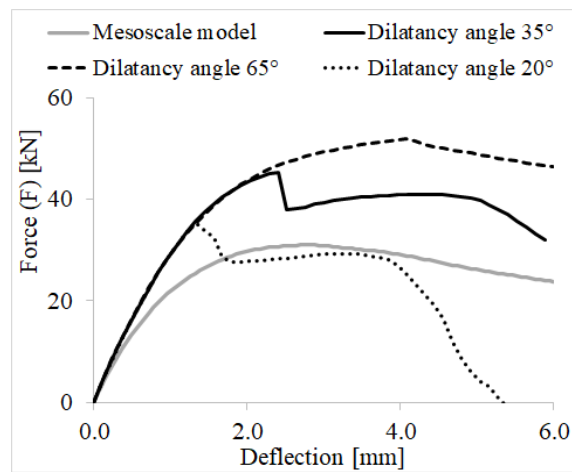

Figure 10. Influence of dilatancy angle on the global response.

The deformed shapes at failure (at the last step of the analysis) with the tensile damage contour distributions are depicted in Figure 11. The models with strong masonry exhibit flexural failure, which is in agreement with the failure mode predicted by the mesoscale models (Figure 7c,d). The models with weak masonry indicate local shear failure in the arch with a marked punching shear effect developing underneath the area where the external load is applied (Figure 11a,b). This is not predicted by the mesoscale model, which shows ring separation at failure (Figure 7a,b). This main difference confirms the inability of the continuum isotropic damage-plasticity model to represent shear sliding between adjacent rings, which is a characteristic failure mechanism of multi-ring arches well captured by detailed mesoscale models. Moreover for the arches with strong masonry, the use of the macroscale continuum isotropic model leads to a significant overestimation of the ultimate strength and ductility, where the numerical predictions are affected significantly by some model parameters (e.g. $w_c$ and $\psi$) which cannot be determined via simplified calibrations.

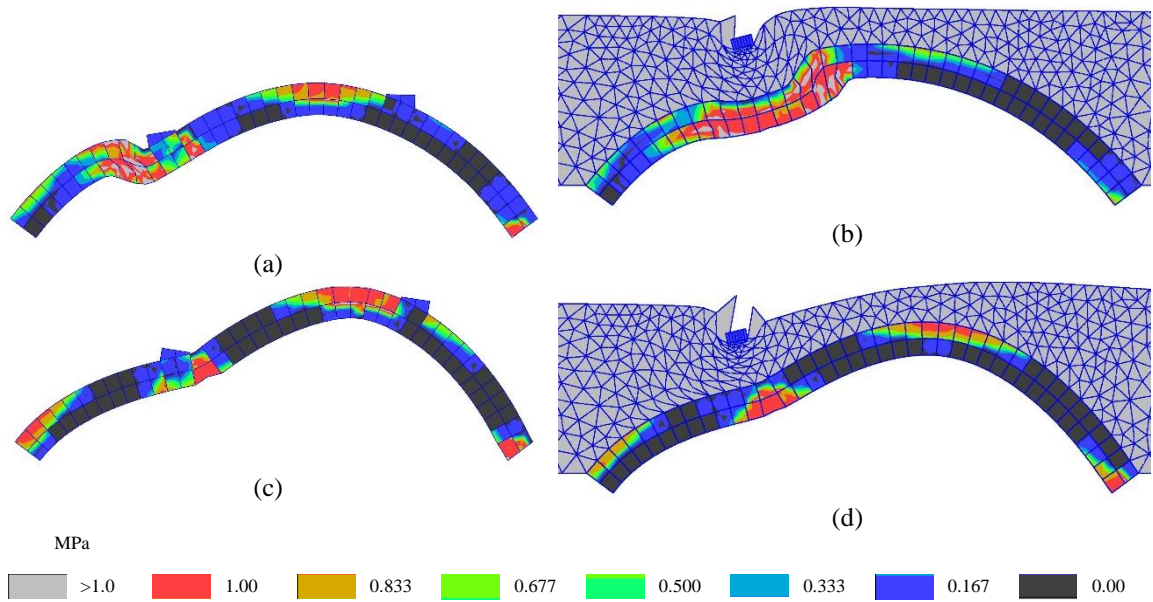

Figure 11.   Deformed shape and damage in tension contours at failure for the weak bare (a) and confined (b) arches; and for strong bare (c) and confined (d) arch specimens.

*4.3.2. Advanced calibration procedure*

To improve the accuracy of the macroscale predictions, the advanced calibration procedure described in Section 3 has been applied to determine the macromodel material parameters, focusing on the specimens with weak masonry where the initial macroscale predictions, based on a simplified calibration of the model material parameters, were not in good agreement with the mesoscale results.

The bare arch in Figure 5a, subjected to two constant initial forces at the quarter span (L/4) and three-quarter span (3/4L), both equal to 16kN, and to a patch load applied at L/4 and increased up to collapse, is used as the virtual test for the calibration of the model parameters. The specific loading condition ignoring the self-weight contribution of masonry has been chosen to activate both flexural damage and shear sliding between adjacent the rings, thus providing suitable information to the optimisation algorithm. It should be pointed out that the selection of appropriate virtual tests which should activate the most critical failure modes of the investigated masonry specimens is the critical step for a successful application of the proposed calibration strategy.

Six model parameters are considered as unknowns in the optimisation procedure: the Young modulus ($E$), the tensile strength ($f_t$) and fracture energy ($G_t$), the ratio ($\tilde{f}_y$) between the yielding and ultimate compression strength, the parameter governing stiffness recovery from tension to compression ($w_c$, see Section 2.1) and the angle of dilatancy ($\psi$). The range for each parameter is reported in the Table 4. The remaining parameters of the solid element are fixed equal to the default values in Table 3.

A load-based partitioning strategy is used for the solution of the calibration problem based on two objectives as defined in Eqs (2), (4) with $\omega_1$, $\omega_2$ the errors due to the loads at L/4 (*F₁*) and 3/4L (*F₂*), normalised with respect to a reference value with the same units (final squared strain energy, divided by the time interval of the virtual test, [$J^2/s$]).

The evaluated PF appears discontinuous; the minimum of $\omega_2$ ($4 \cdot 10^{-3}$) is reached for $\omega_1 = 0.10$. Conversely, the minimum of $\omega_1$ ($6 \cdot 10^{-4}$) corresponds to a much larger value $\omega_2 = 1.19$. This implies that calibrating the response on the force-displacement curve of the variable load at L/4 (minimum $\omega_1$) may entail significant error in the total energy.

In Figure 12a, the solutions on the PF are shown using the normalised errors $\omega_1^* = (\omega_1 - \omega_1^{min})/(\omega_1^{max} - \omega_1^{min})$ and $\omega_2^* = (\omega_2 - \omega_2^{min})/(\omega_2^{max} - \omega_2^{min})$ ranging from 0 to 1. In the following, the solution corresponding to the minimum global error $\omega_{min}^* = min\left\{\sqrt{\omega_1^{*2} + \omega_2^{*2}}\right\} = 0.078$, highlighted in the graph, is considered as the reference solution. Figure 12b shows the displacement $d_1$ at L/4 against the load $F_1$ associated with all the solutions of the PF compared to the response of the mesoscale model. Three families of curves can be observed: the curves that minimise the error related to $F_1$, which fit very well the response of the mesoscale; the curves that minimise $\omega_2$, i.e., the error related to $F_2$, which provide a significant overestimation of the peak-strength of the arch (from 35 kN to 45 kN) and the curves that minimise the global dimensionless error allowing for the two objectives of the optimisation procedure. These solutions and in particular the solution associated with the minimum error provide a good prediction in terms of initial stiffness and peak load, but show some differences regarding the post-peak stage. In Figure 12c, the vertical displacement at three-quarter $d_2$ span is plotted against the load $F_1$. It is possible to see that in no case the final uplift shown by the mesoscale model is attained by the macroscale models of the PF. However, the dashed line, identifying the compromise solution of the multi-objective optimisation, shows a general good agreement with the mesoscale curve.

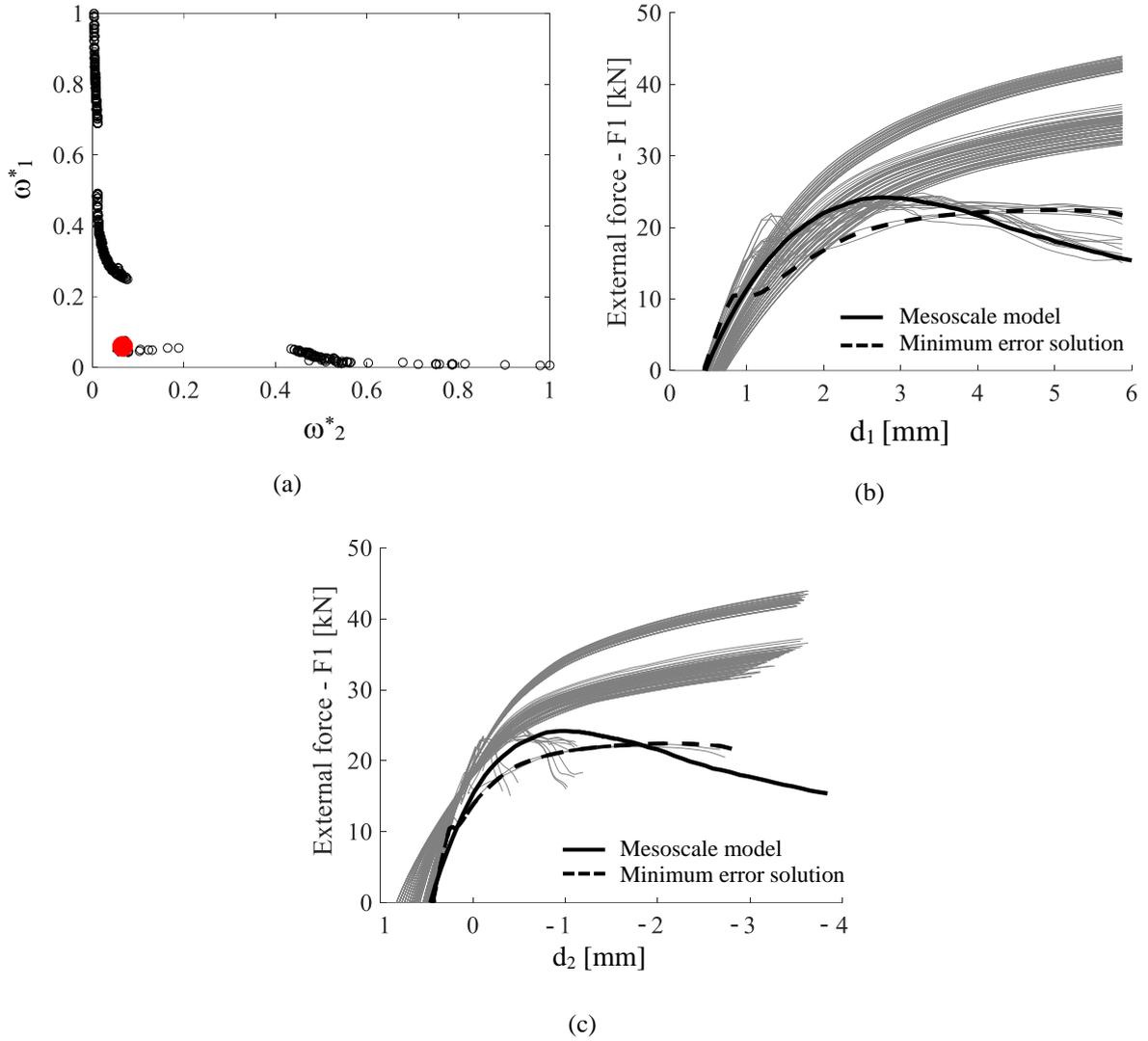

Figure 12. Calibration of the continuum weak model: Pareto Front solutions (a); load-displacement curve at L/4 (b) and 3/4L (c).

In Table 4, the calibrated parameters for the minimum error solution are shown, together with the range identified by the solutions in its surroundings ($\omega^* < 2.5\omega^*_{min}$). These latter values allow investigating the sensitivity of the calibrated response to each parameter, compared to the initial variation range. Analysing the results, all parameters seem rather univocally determined as a significant reduction of the ranges is observed, with the only exception of $\tilde{f}_y$. It is interesting to see that the calibrated $w_c$ is close to zero, unlikely the typical assumption considered in the simplified calibration.

**Table 4**: Input parameters and results of the calibration procedure for the continuum model (weak masonry)

| Parameter | Unit | Initial range | | Solutions with $\omega^* \leq 2.5\omega^*_{min}$ | | Minimum error Solution |
|---|---|---|---|---|---|---|
| | | Lower bound | Upper bound | Upper bound | Upper bound | |
| $E$ | MPa | 1000 | 6000 | 2050 | 2550 | 2500 |
| $\psi$ | º | 0 | 90 | 18 | 26 | 25 |
| $f_t$ | N/mm | 0.01 | 1.0 | 0.024 | 0.041 | 0.025 |
| $G_t$ | N/mm | 0.001 | 0.5 | 0.021 | 0.028 | 0.022 |
| $\tilde{f}_y$ | - | 0.01 | 1.0 | 0.40 | 0.83 | 0.72 |
| $w_c$ | - | 0.0 | 1.0 | 0.00 | 0.16 | 0.02 |

The results of the model calibration are validated by analysing the confined and bare arches subjected to the loading condition described in the initial mesoscale simulation in Section 4.2. Moreover, two additional load conditions for the bare arch are investigated in which the arch is subjected to a concentrated force alternatively applied at the mid span and at one eight span without initial symmetric forces. The load-displacement curves for the weak masonry models are shown in Figure 13, where the continuum calibrated model is compared against the mesoscale model and the continuum model with simplified calibration procedure for $w_c$=0. Generally, the model calibrated by the advanced procedure exhibits a much-improved agreement with the mesoscale model. However, in the case of the bare arch loaded at the quarter-span, the continuum model shows a premature shear failure which leads to a significant underestimation of the maximum load and displacement capacity of the arch (Figure 13c). In the case of bare arch loaded at the mid span (Figure 13a), the macromodel calibrated by the advanced procedure underestimates the peak-load value. It provides, however, an adequate prediction of the residual strength and pre/post peak response. Finally, a satisfactory comparison can be observed in the cases of bare arch loaded at one eighth span and the confined arch specimen (Figure 13b and 13d).

The failure mechanism of the bare arch, displayed in Figure 14a, is characterised by an evident punching effect due to the shear failure of the element of the arch underneath the load, which is not observed in Figure 7b. On the contrary, the failure mechanism of the confined arch (Figure 14b) is rather consistent with the mechanism obtained by the mesoscale model (Figure 7d). The sliding between adjacent rings is represented by shear failure of the solid elements; however, unlikely the continuum model calibrated by means of the simplified procedure (Figure 11b), the punching effect is not observed.

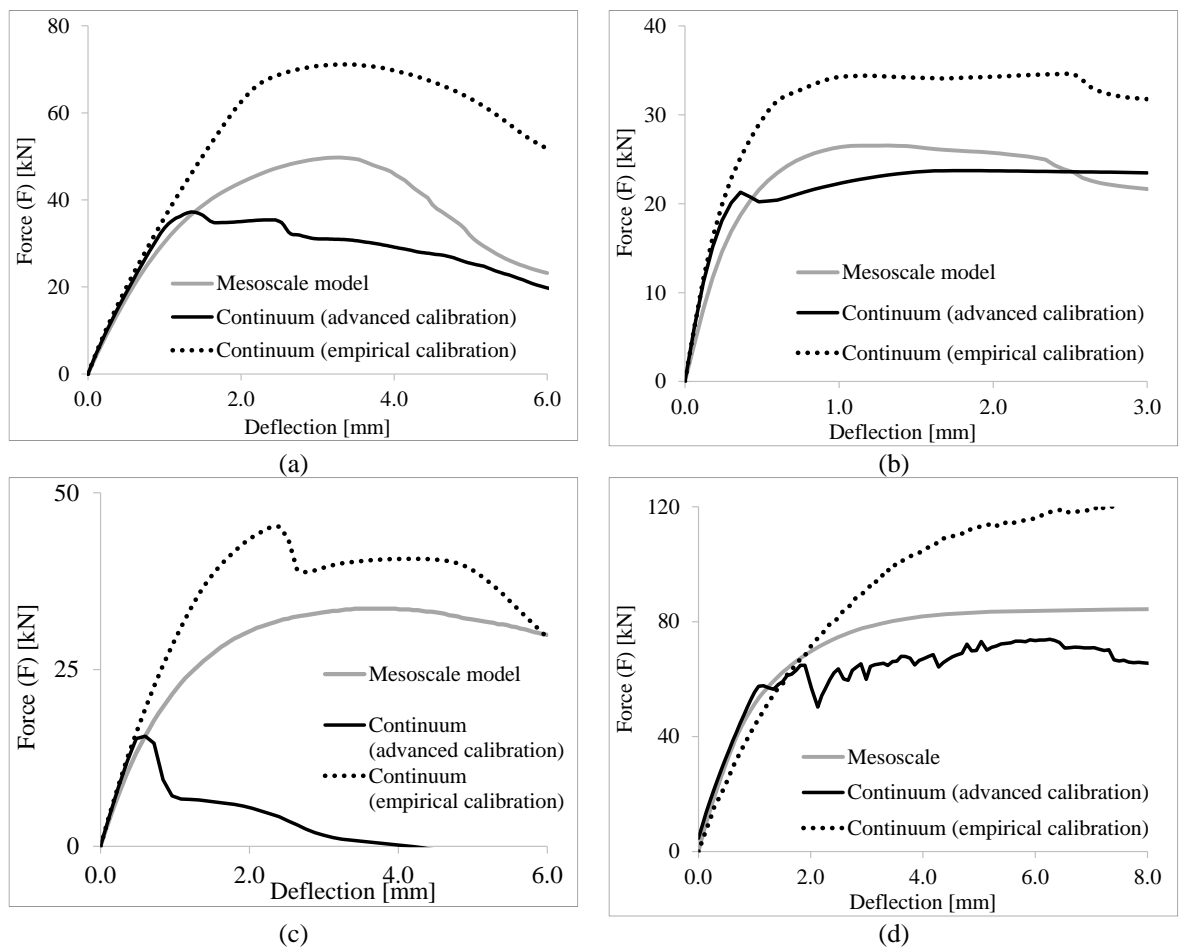

Figure 13. Capacity curve of the bare arch loaded with a concentrated force at (a) the middle span, (b) 1/8th of span and (c) quarter span with initial forces; (d) confined arch.

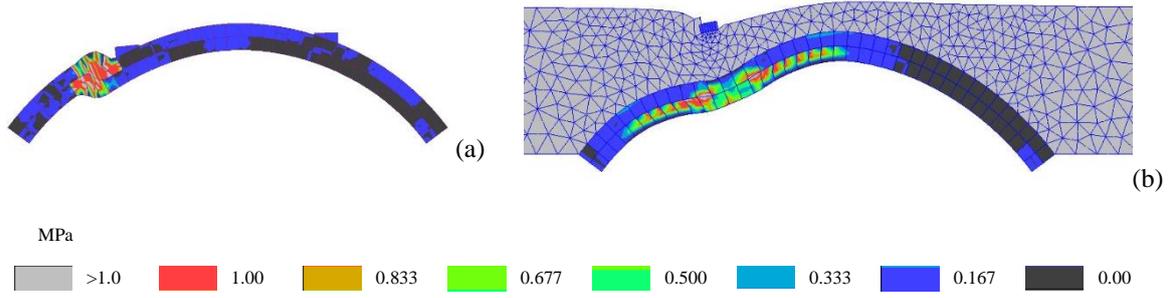

Figure 14. Continuum model calibrated: failure mechanism and distribution of damage index in tension of the (a) bare arch and the (b) confined arch.

In conclusion, it can be affirmed that the rigorous calibration procedure allows for a substantial improvement of the continuum model predictions in representing the confined arch specimen. However, the calibrated continuum model still shows evident limits in simulating the bare arch response, as it provides an unrealistic failure mode underestimating both the global strength and ductility of the arch.

It is worth pointing out that despite the fact that the adopted calibration strategy is based on the use of the entire arch specimen for the virtual test, its applicability is still computationally efficient as it applied on a simple 2D strip model neglecting the backfill and its interaction with the arch.

### 4.4 Hybrid model simulations

In the final study, the proposed hybrid model described in Section 2 is employed to simulate the bare and the confined arch of Figure 5. The model material parameters are calibrated through the rigorous procedure described in Section 3 and applied before to determine the material properties for the continuum macroscale model. Nine model parameters are calibrated by the optimisation procedure. Namely, the same six parameters of for the solid elements, already considered in Section 4.3.2 with the addition of three parameters characterising the interfaces: the shear stiffness ($k_t^M$), the cohesion ($c^M$) and the shear fracture energy ($G_s^M$). The remaining interface parameters are assumed either coincident to the parameters of the mesoscale model ($f_c, G_c, tg\phi, tg\phi_g$) or proportional to the parameters assumed as unknown in the optimisation ($k_n^M, f_t^M, G_t^M$) as indicated below:

$$k_t^M = k_t^M \cdot \frac{k_n^m}{k_t^m}$$

$$f_t^M = c^M \cdot \frac{f_t^m}{c^m} \qquad (5)$$

$$G_t^M = G_s^M \cdot \frac{G_t^m}{G_s^m}$$

where the superscripts $M$ and $m$ refer to the macroscale and mesoscale representation, respectively.

In the following, the interface stiffness and cohesion parameters are represented by the non-dimensional coefficients:

- $k^* = 2k_t^M t(1+\nu)/E$, where $E$ and $\nu$ are the Young modulus and the Poisson's coefficient of the solid elements and $t$ a fictitious thickness equal to 1 mm.

- $c^* = c^M/f_{vo}$, where $f_{vo}$ is the initial shear strength of the solid elements (Eq. (5) with $\kappa = 0$).

The variation ranges of the unknown parameters are reported in Table 5. The calibration was performed for both weak and strong masonry by using the same procedures and objective functions as for the calibration of the continuum model.

Figure 15a displays the solutions belonging to the PF ($\omega_1^*$ - $\omega_2^*$) for the calibration of the weak masonry model, while the corresponding load-displacement curves are reported in Figure 15b,c. Comparing these results to the solutions of the optimisation for the continuum model in Section 5.1, it can be observed that:

- The solutions of the PF are uniformly distributed in the space $\omega_1$-$\omega_2$ and are associated with much lower errors.

- The load-displacement curves are less dispersed and much closer to the mesoscale predictions.

These remarks confirm that the further free parameters included in the optimisation algorithm ($k^*, c^*, G_t^M$) effectively improve the quality of the results. In this case, the absolute minimum of $\omega_2$ ($4.88 \cdot 10^{-3}$) is reached for a value of $\omega_1 = 4.97 \cdot 10^{-3}$ while the minimum of $\omega_1$ ($4.66 \cdot 10^{-5}$) corresponds to $\omega_2 = 7.57 \cdot 10^{-3}$. Following the procedure in Section 5.1, the two errors are normalised leading to a minimum solution error $\omega_{min}^* = min\left(\sqrt{\omega_1^{*2} + \omega_2^{*2}}\right) = 0.2284$ which corresponds to the set of model parameters reported in Table 5.

The PF for the model with strong masonry is shown in Figure 16a. The minimum error of $\omega_2 = 0.011$ is reached for $\omega_1 = 0.0056$, while the minimum error $\omega_1 = 0.0012$ is associated with $\omega_2 = 0.337$. The minimum solution error corresponds to $\omega_{min}^* = 0.3207$, (A in Figure 16a) and the matching set of parameters are indicated in Table 5. It can be noticed that the latter solution is characterised by a low value of the dimensionless cohesion ($c^*$=0.69). This circumstance may potentially lead to a response characterised by sliding between the ring which is in disagreement with the mesoscale results. For this reason, another solution (B in Figure 16a), corresponding to $\omega_B^* = 0.4673$, is also considered.

The response curves of the PF solutions are shown in Figure 16b and 16c with the two reference solutions reported with dashed lines. Two families of curves, which tend to minimise the two objectives separately are visible. Although the PF is less regular than the previous case, the selected solutions confirm a good match with the mesoscale curves.

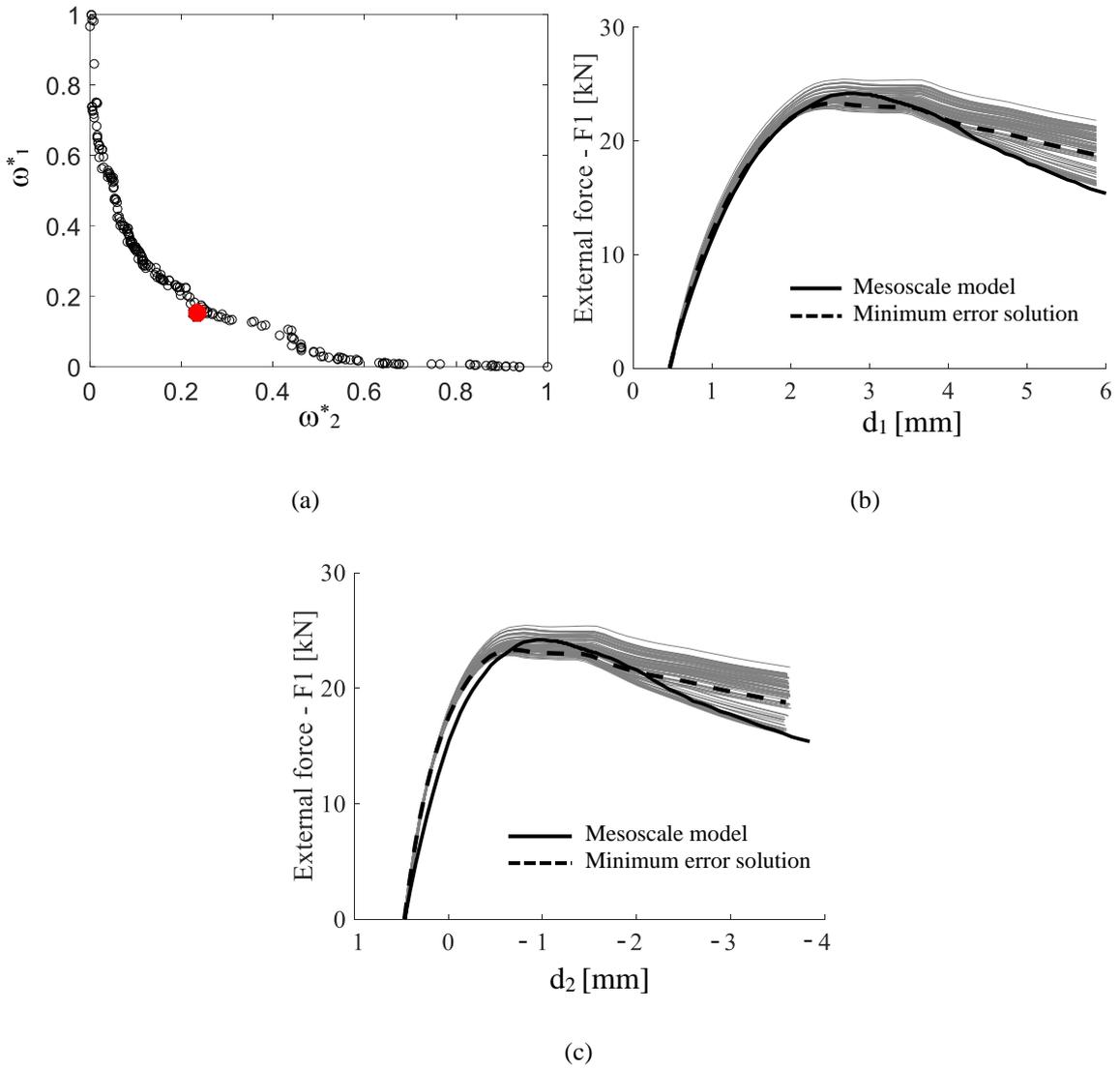

Figure 15. Calibration weak model: Pareto Front solutions (a); capacity curves at L/4f (b) and 3/4L (c).

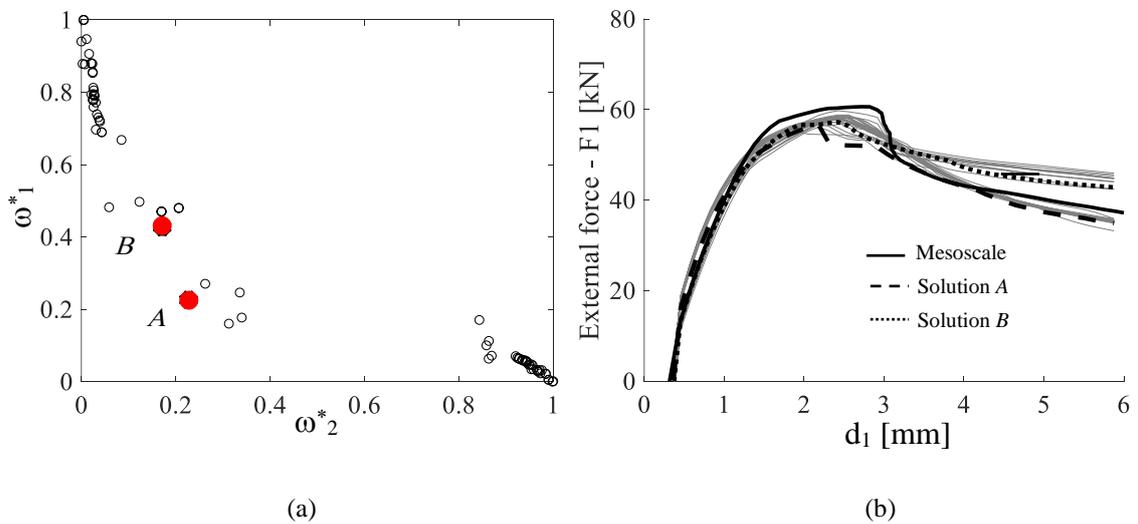

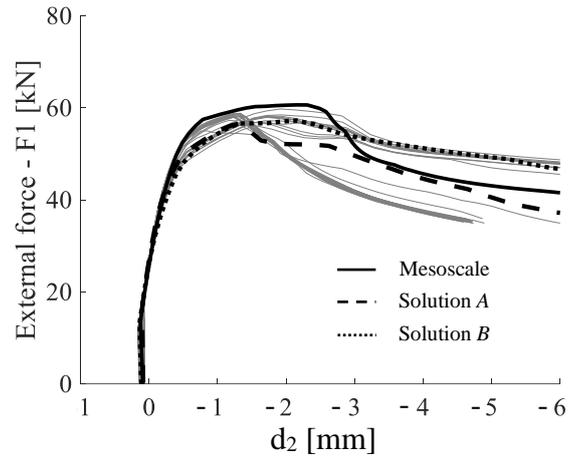

(c)

Figure 16.    Calibration Strong model: Pareto Front solutions (a); capacity curves at L/4f (b) and 3/4L (c)

Table 5: Input parameters and results of the calibration procedures for the hybrid model.

| Parameter | Unit | Initial range | | Minimum error solution weak masonry | Minimum error solutions strong masonry | |
|---|---|---|---|---|---|---|
| | | Lower bound | Upper bound | | Solution A | Solution B |
| $E$ | MPa | 1000 | 6000 | 2350 | 4700 | 4200 |
| $\psi$ | ° | 0 | 90 | 63 | 23 | 59 |
| $f_t$ | N/mm | 0.01 | 1.0 | 0.083 | 0.164 | 0.137 |
| $G_t$ | N/mm | 0.001 | 0.5 | 0.089 | 0.042 | 0.03 |
| $\tilde{f}_y$ | - | 0.01 | 1.0 | 0.79 | 0.71 | 0.68 |
| $w_c$ | - | 0.0 | 1.0 | 0.10 | 0.48 | 0.00 |
| $k^*$ | - | 0.01 | 1.5 | 0.01 | 0.01 | 0.04 |
| $c^*$ | - | 0.0 | 2.0 | 0.62 | 0.69 | 1.44 |
| $G_t^M$ | N/mm | 0.01 | 0.25 | 0.096 | 0.167 | 0.213 |

Analogously to the procedure for the continuum model in Section 4.3.2, the results of the calibration are validated considering the bare and confined specimens, plus two further models representing the bare arch subjected to a concentrated force at mid span and at one-eighth span. The load-displacement curves of the hybrid model are displayed in Figure 17, where they are compared against the mesoscale curves and the predictions of the continuum macroscale model calibrated by the advanced procedure in Section 4.3.2.

The results of the hybrid macromodel are in a good agreement with those obtained by the mesoscale model confirming a generally improved prediction compared to the results provided by the continuum macromodel. The only exception is represented by the load condition with the force at L/8, where the continuum model provides a better prediction of the peak-loads. However, the curve of the hybrid model, also in this case, is more consistent to the mesoscale curve in the pre- and post-peak stages. It appears that the presence of the backfill reduces the differences between the results, with mesoscale and hybrid model curves almost coincident.

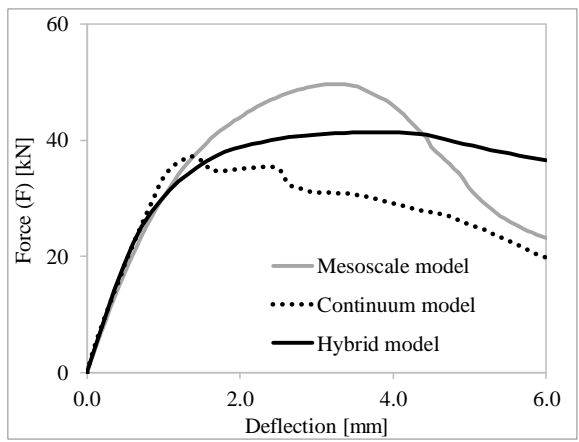
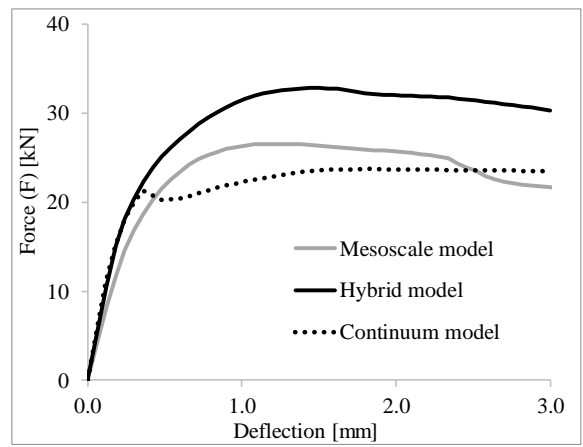

(a)                                  (b)

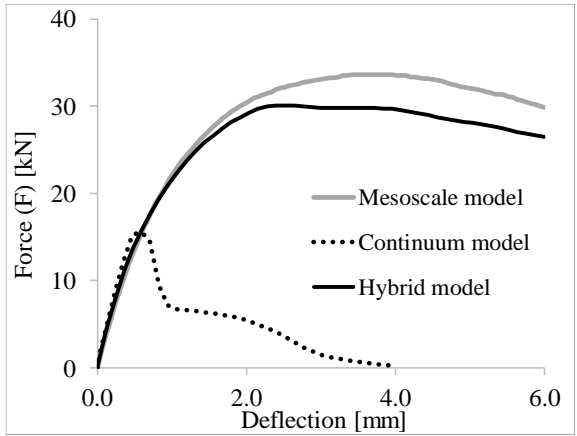 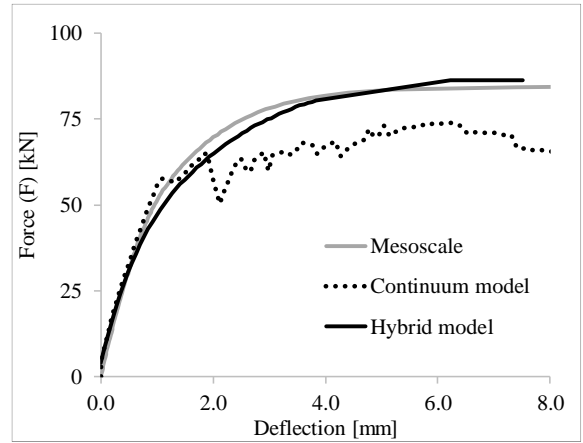

|(c)|(d)|
|---|---|

Figure 17. Weak model: capacity curve of the bare arch loaded with a concentrated force at (a) the middle span, (b) 1/8th of span and (c) quarter span with initial F0 forces; (d) arch interacting with backfill.

The failure modes of the bare arch and the arch interacting with backfill are displayed in Figure 18, where the von-Mises equivalent stress distribution in the solid elements and the damage contours on the interface elements are also shown. A good agreement between the failure modes of the hybrid model and those obtained by the mesoscale model, both in terms of stress distribution (Figure 6) and damage index distribution along the ring-to-ring and arch-to-backfill interfaces (Figure 7). Importantly, the ring sliding mechanism, which could not be predicted by the continuum model, is well described by the proposed hybrid macroscale representation.

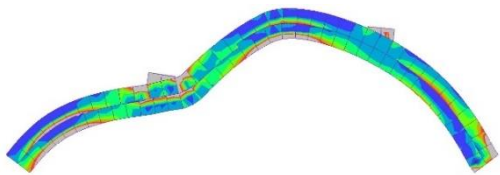 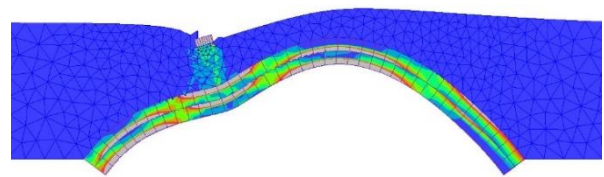

|(a1)|(b1)|
|---|---|

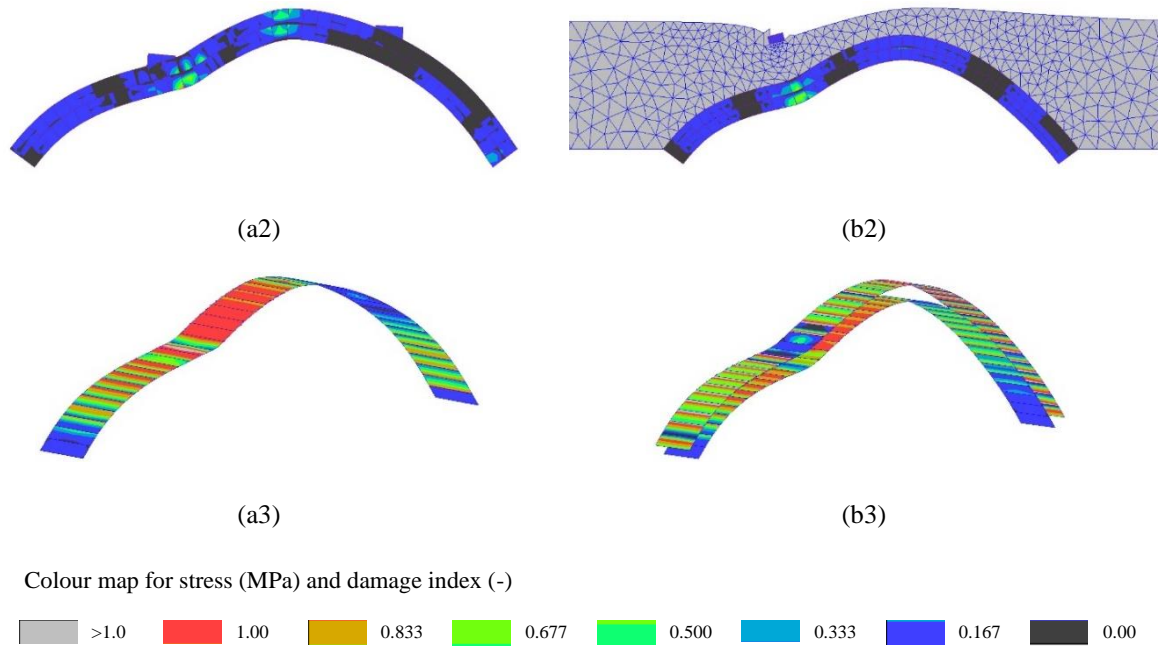

Colour map for stress (MPa) and damage index (-)

| >1.0 | 1.00 | 0.833 | 0.677 | 0.500 | 0.333 | 0.167 | 0.00 |

Figure 18.  Weak model: Ultimate state of (a) the bare arch and (b) the arch interacting with backfill subjected to a force at the span quarter: Von-Mises stress (a1, b1), damage index in tension of the solid elements (a2, b2); damage index of the interfaces (a3, b3).

The results of the calibration analyses in term of load-displacement curves and failure mechanisms for the strong masonry model are shown in Figure 19 and in Figure 20, respectively. The curves obtained using the solution *B* parameters indicate a good agreement with the mesoscale results for all the considered models and loading conditions. Conversely, solution *A* leads to a significant overestimation of the ultimate strength of the structure in the cases of the bare arch loaded at L/2 (Figure 19a) and the arch interacting with backfill (Figure 19d). The failure mechanisms obtained by the macromodel calibrated with the solution *B* (Figure 20) result in a good agreement with those obtained by the mesoscale model (Figure 7). In conclusion, it can be stated that the advanced calibration procedure leads to a realistic set of mechanical parameters to describe the global response of the arches, including strong masonry arches, under a wide range of boundary and loading conditions. The adopted approach to select the reference solution from the results of the multi-objective optimization procedure, based on the analysis of the Pareto Front, appears to be sufficiently accurate and robust. However, the circumstance by which the minimum error solution (*A*) provided less satisfactory results compared to those obtained by another solution belonging to the PF solution (solution *B*) denotes that further improvements to the selection strategy from the PF and/or the definition of the multi-objective optimization problem may be needed. This open issue will be investigated by the authors in future studies.

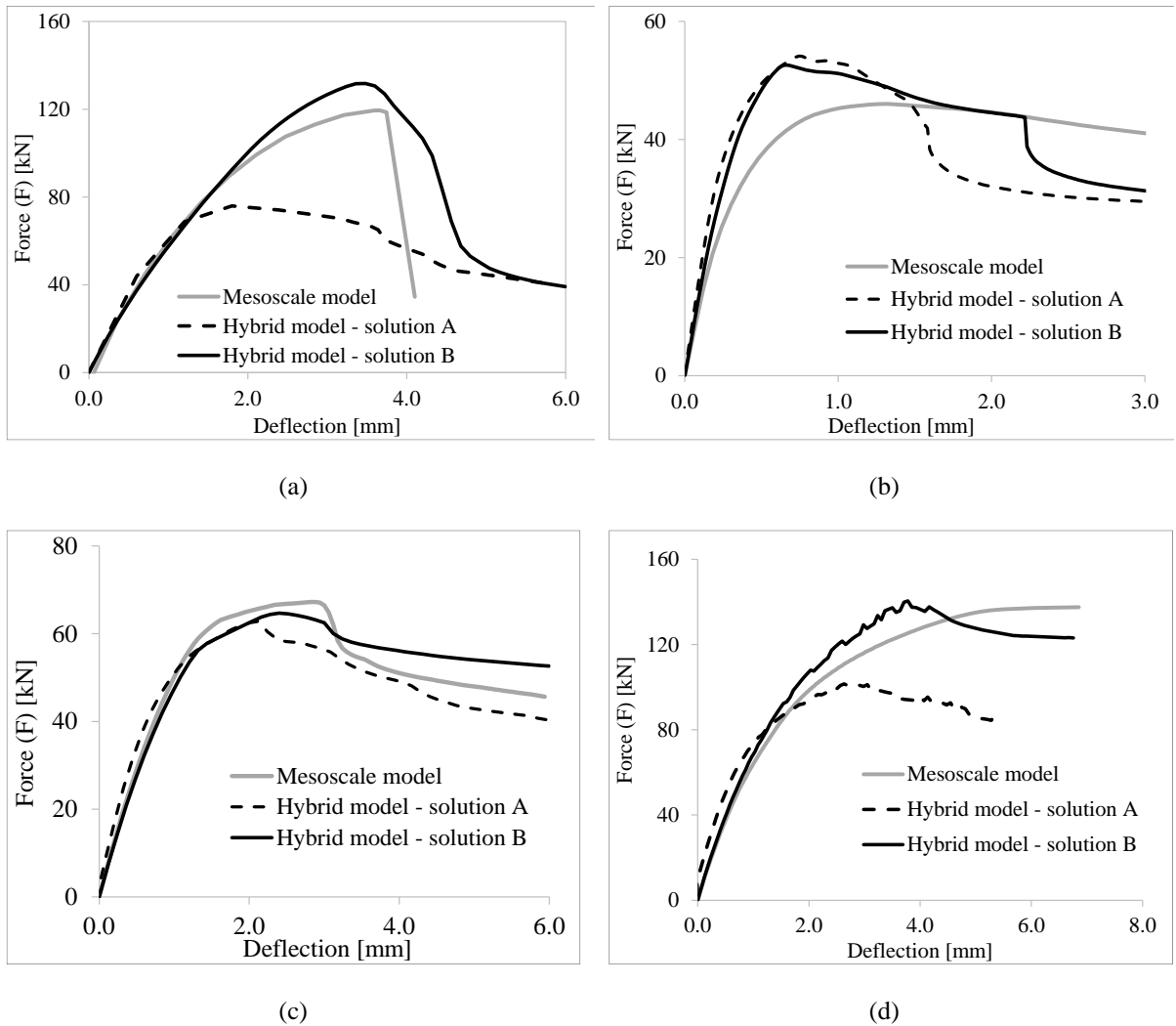

Figure 19. Load-displacement curves for the strong masonry bare arch with *F* at (a) mid-span, (b) one eight span, (c) quarter span and (d) for the arch interacting with backfill loaded at quarter span.

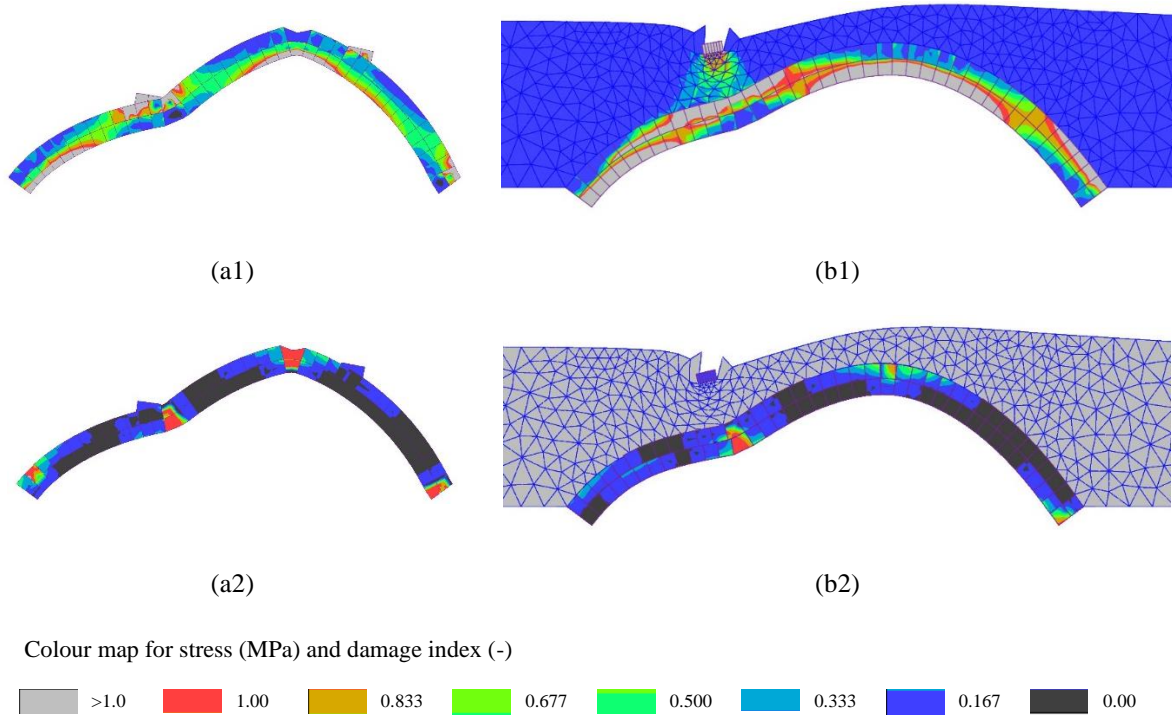

Figure 20. Weak model: Ultimate state of (a) the bare arch and (b) the arch interacting with backfill subjected to a force at the span quarter: Von-Mises stress (a1, b1), damage index in traction of the solid elements (a2, b2).

## 5 CONCLUSIONS

In this study, a hybrid continuum-discrete macro-modelling description for brick-masonry multi-ring arches and arch bridges is proposed. According to this modelling approach, the arch and backfill domain are modelled by 3D continuum solid elements while 2D zero-thickness nonlinear interfaces, arranged along the mid-thickness and extrados circumferential surfaces, simulate the potential ring separation and the interaction between the arch and backfill. Two advanced damage-plasticity constitutive models are employed for the 3D solid and interface elements. An effective and robust multi-level calibration procedure, based on minimisation of stress power error solved by means of Genetic Algorithms, is enhanced by partitioning the error functional and used to evaluate the model mechanical parameters employing the results from detailed mesoscale simulations on virtual experiments.

The accuracy and potential of the proposed modelling strategy and calibration procedure is demonstrated by analysing 2D-strip masonry arch specimens, also interacting with backfill, characterised by different failure mechanisms.

The proposed hybrid continuum-discrete macroscale model describes the masonry arch specimens by approximately 5% of DoFs required by the mesoscale description. This feature leads to enhanced computational efficiency of the proposed model, which may be crucial in simulating large arch bridges.

The results of the proposed hybrid model are compared to those obtained by detailed mesoscale and continuum finite element macroscale descriptions. The obtained results confirm the ability of the proposed modelling strategy to predict the ultimate strength and displacement capacity of multi-ring masonry arches and the corresponding failure mechanisms, allowing for the presence of sliding between the rings. On the other hand, the numerical results identified some drawbacks associated with the use of conventional isotropic finite element macromodels, which can be summarised as following:

- The use of continuum finite element macromodels, without a rigorous calibration of the mechanical parameters, can lead to an inaccurate and non-objective prediction of the arch response.
- Continuum finite element macromodels, even when calibrated by means of rigorous procedures, present some drawbacks in simulating the ultimate arch behaviour when it is driven by sliding between adjacent rings.

Both these limitations may significantly affect the results of safety assessments of masonry arch bridges, leading to a crudely approximated or completely misleading prediction of the effective safety level of the bridge and its mode of failure. In this regard, the proposed hybrid modelling strategy offers the possibility to significantly improve the accuracy of the numerical predictions without requiring a significant increase in the computational effort associated with the nonlinear analysis.

ACKNOWLEDGEMENTS

The first author gratefully acknowledges support from the Marie Skłodowska-Curie Individual fellowship under Grant Agreement 846061 (Project Title: Realistic Assessment of Historical Masonry Bridges under Extreme Environmental Actions, ''RAMBEA'', https://cordis.europa.eu/project/id/846061.